\documentclass[a4paper]{jpconf}

\usepackage[english]{babel} 

\usepackage[%
	final,%
]{graphicx}
\usepackage{overpic}

\usepackage[
   centertags, 
   sumlimits,  
   intlimits,  
   namelimits, 
]{amsmath} %
\usepackage{amssymb}
\newcommand{\unit}[1]{\, {\rm #1}}
\usepackage[%
	pdfstartview=FitH,
	pdfpagemode=UseNone,
	bookmarksopen=true
	]{hyperref}

\usepackage[all]{hypcap} 

\usepackage{microtype} 

\makeatletter
\def\tagform@#1{\maketag@@@{\ignorespaces#1\unskip\@@italiccorr}}
\let\orgtheequation\theequation
\def\theequation{(\orgtheequation)}
\makeatother

\usepackage{xspace}

\newcommand{\beq}{\begin{equation}}
\newcommand{\eeq}{\end{equation}}

\begin{document}

\title{Heavy vs.\ light flavor energy loss within a partonic transport model }
\author{Jan Uphoff,$^1$ Oliver Fochler,$^1$ Zhe Xu,$^{2}$ and Carsten Greiner$^1$}

\address{$^1$ Institut f\"ur Theoretische Physik, Johann Wolfgang 
Goethe-Universit\"at Frankfurt, Max-von-Laue-Str. 1, 
D-60438 Frankfurt am Main, Germany}
\address{$^2$ Department of Physics, Tsinghua University, Beijing 100084, China}

\ead{uphoff@th.physik.uni-frankfurt.de}

\begin{abstract}
The full space-time evolution of gluons, light and heavy quarks in ultra-relativistic heavy-ion collisions is studied within the partonic transport model \emph{Boltzmann Approach to MultiParton Scatterings} (BAMPS). We discuss for light and heavy quarks the elastic and radiative energy loss with a running coupling. Radiative processes, in particular, are implemented through an improved version of the Gunion-Bertsch matrix element, which is derived from comparisons to the exact result, explicitly taking finite heavy quark masses into account. Subsequently, we present results with and without radiative processes and compare them to experimental data at LHC. A focus is put on the nuclear modification factor of charged hadrons and $D$ mesons.
\end{abstract}

\section{Introduction}

In ultra-relativistic heavy-ion collisions at the Relativistic Heavy Ion Collider (RHIC) at BNL \cite{Adams:2005dq,Adcox:2004mh} and the Large Hadron Collider (LHC) at CERN \cite{Muller:2012zq} a hot and dense medium is produced. Since quarks and gluons form the relevant degrees of freedom, it is commonly named quark gluon plasma (QGP). Interesting properties are, for instance, a collective behavior like a nearly perfect fluid or the quenching of highly energetic particles.

In particular charm and bottom quarks provide a unique way to gain insight in the properties of this matter. Since they are heavy, their production time is at a very early stage of the heavy-ion collision when enough energy is available \cite{Uphoff:2010sh}. Therefore, they traverse the medium right from the beginning for a rather long time, interact with other medium particles, participate in the flow, and lose energy. 
The mechanism responsible for their strong interaction with the medium is actively debated, most recently, for instance, in
\cite{Abir:2012pu,Meistrenko:2012ju,Uphoff:2012gb,He:2012df,Horowitz:2012cf,Buzzatti:2012pe,Cao:2012au, Nahrgang:2013saa,Alberico:2013bza,Mazumder:2013oaa,Lang:2013cca}.

In this paper we study the nuclear modification factor and elliptic flow of heavy flavor particles at LHC within the partonic transport model BAMPS. Furthermore, we calculate the elastic and radiative energy loss of light and heavy quarks in a static thermal medium. To this end, we also compare the $D$ meson nuclear modification factor to that of charged hadrons obtained with elastic and radiative processes.

\section{Partonic transport model BAMPS}
BAMPS (\emph{Boltzmann Approach to MultiParton Scatterings}) \cite{Xu:2004mz,Xu:2007aa} is a partonic transport model  that describes the full space-time evolution of the QGP by solving the Boltzmann equation,
\begin{equation}
\label{boltzmann}
\left ( \frac{\partial}{\partial t} + \frac{{\mathbf p}_i}{E_i}
\frac{\partial}{\partial {\mathbf r}} \right )\,
f_i({\mathbf r}, {\mathbf p}_i, t) = {\cal C}_i^{2\rightarrow 2} + {\cal C}_i^{2\leftrightarrow 3}+ \ldots  \ ,
\end{equation}
for on-shell partons and pQCD interactions. 
For light partons, all possible $2\rightarrow 2$ and $2 \leftrightarrow 3$ processes are included. On the heavy flavor sector, all relevant elastic and radiative collisions of heavy quarks with other medium constituents have been implemented in BAMPS, while the running  of the coupling is explicitly taken into account for elastic as well as radiative light and heavy flavor processes. 
The divergent $t$ channel of heavy flavor scatterings is regularized with a screening mass $\mu$ that is determined by matching elastic energy loss calculations with leading order pQCD cross sections to results from hard thermal loop (HTL) calculations. The comparison of both results shows that the screening mass~$\mu$ is smaller than the usually employed Debye mass~$m_D$, more precisely, $\mu^2 = \kappa m_D^2$ with $\kappa = 1/(2e) \approx 0.2$ \cite{Gossiaux:2008jv,Peshier:2008bg,Uphoff:2011ad}.

For radiative processes we generalize the recently calculated improvement of the \textsc{Gunion} and \textsc{Bertsch} (GB) cross section \cite{Fochler:2013epa} to finite masses \cite{Uphoff:GB}. The result is consistent with the calculation of Ref.~\cite{Aichelin:2013mra}.
The dead cone suppression \cite{Dokshitzer:2001zm,Abir:2011jb} of small angle radiation due to the finite heavy quark mass is explicitly present in our result.

\section{Results and comparison with data}

\begin{figure}[t]
\begin{minipage}[t]{0.49\textwidth}
\centering
\includegraphics[width=1.0\textwidth]{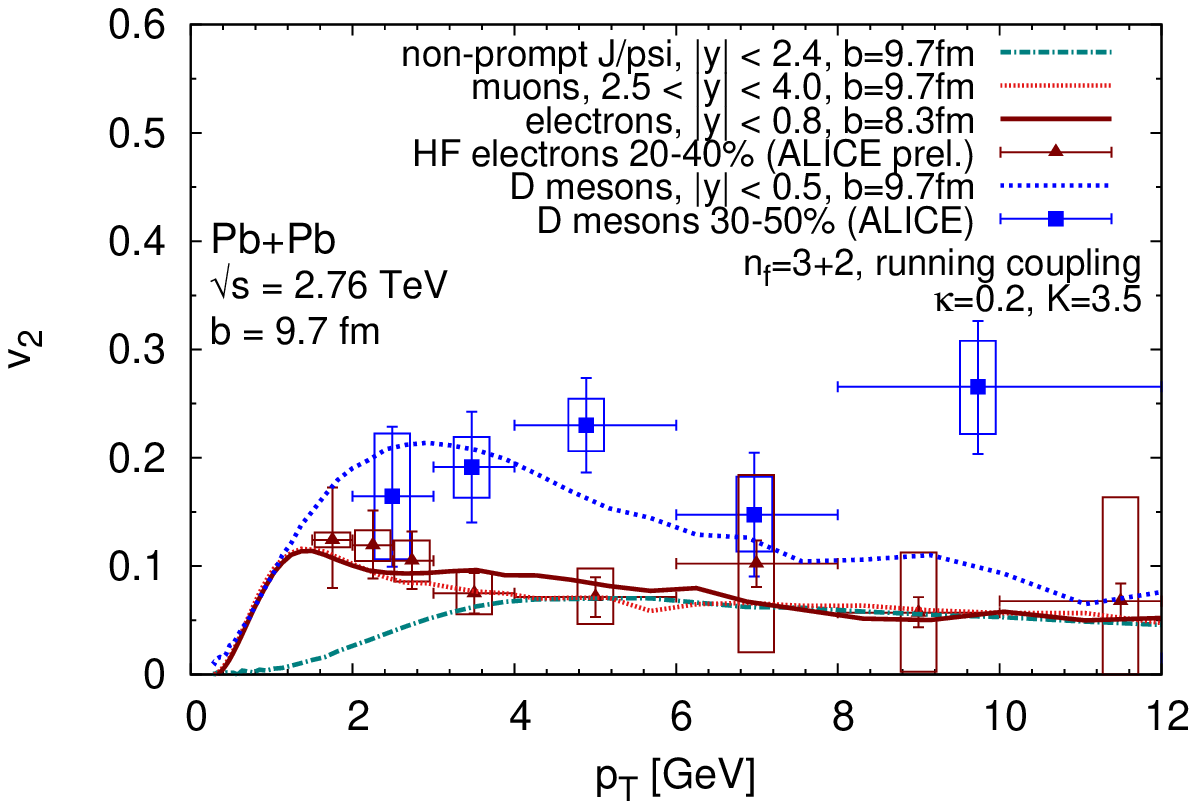}
\end{minipage}
\hfill
\begin{minipage}[t]{0.49\textwidth}
\centering
\includegraphics[width=1.0\textwidth]{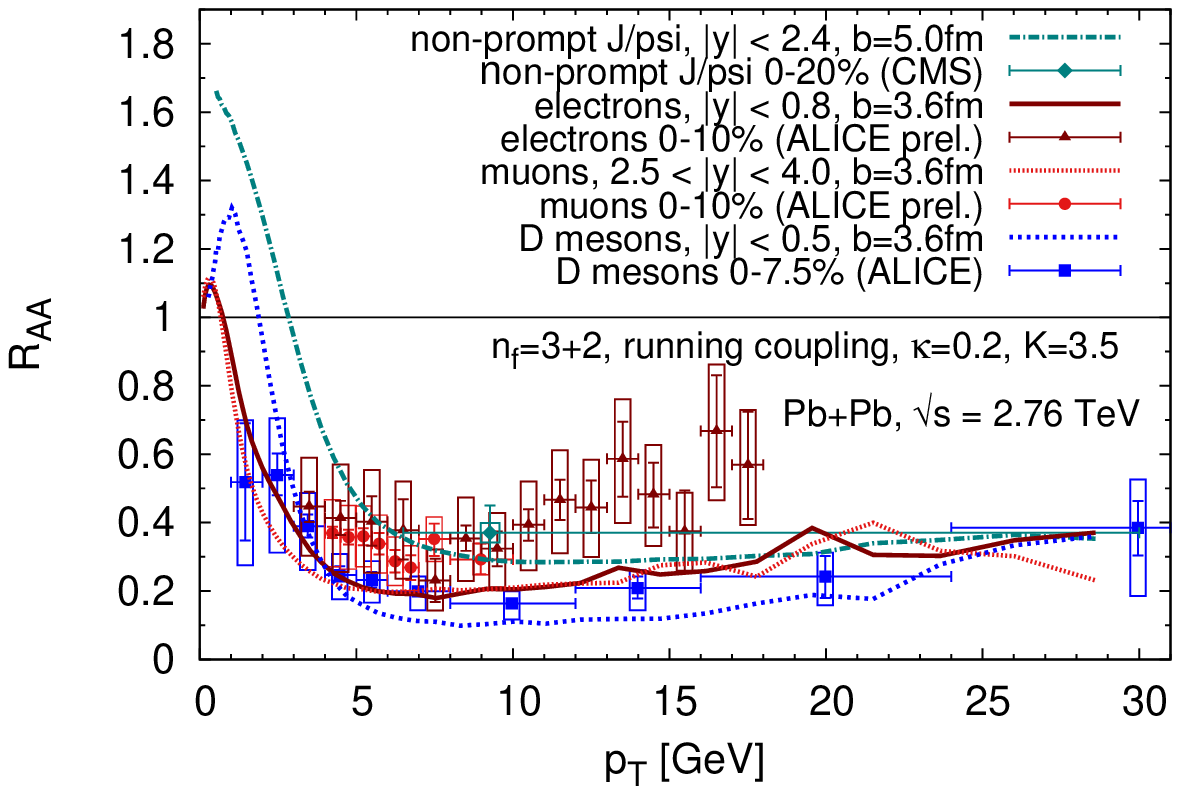}
\end{minipage}
\caption{Elliptic flow $v_2$ (left) and nuclear modification factor $R_{AA}$ (right) of various heavy flavor particles at LHC together with data \cite{Abelev:2013lca,Sakai:2013ata,delValle:2012qw,Abelev:2012qh,Grelli:2012yv,Chatrchyan:2012np}. Only binary heavy flavor processes are considered and multiplied with $K=3.5$.
}
\label{fig:v2_raa_lhc}
\end{figure}

Quantitative BAMPS comparisons \cite{Uphoff:2010bv,Uphoff:2011ad,Fochler:2011en,Uphoff:2012gb} show that elastic processes with a running coupling and the improved screening procedure contribute significantly to the energy loss of heavy quarks. However, they alone cannot reproduce the data of the nuclear modification factor or the elliptic flow of any heavy flavor particle species. 
Before radiative heavy quark processes have been implemented in BAMPS, we mimicked their influence by effectively increasing the elastic cross section by a factor $K=3.5$, which is tuned to  the  elliptic flow data of heavy flavor electrons at RHIC \cite{Uphoff:2012gb}. Simultaneously, the nuclear modification factor  of heavy flavor electrons at RHIC can be described with the same parameter. Having fixed this parameter to the RHIC data, we find a good agreement with the experimentally measured nuclear modification factor and elliptic flow of all heavy flavor particles at LHC (see Fig.~\ref{fig:v2_raa_lhc}). Most of these calculations were predictions  \cite{Uphoff:2012gb}.
%

However, the need of the phenomenological $K$ factor is rather unsatisfying from the theory perspective.
Therefore, the question arises whether radiative processes can account for the missing contribution parameterized by the $K$ factor. To this end, we present in the following preliminary BAMPS calculations including radiative processes for both light and heavy particles.
Since the improved screening prescription is only derived for heavy quarks, we employ in the following the standard Debye screening ($\kappa=1$) to treat light and heavy partons consistently.

\begin{figure}[t]
\begin{minipage}[t]{0.49\textwidth}
\centering
\begin{overpic}[width=1.0\textwidth]{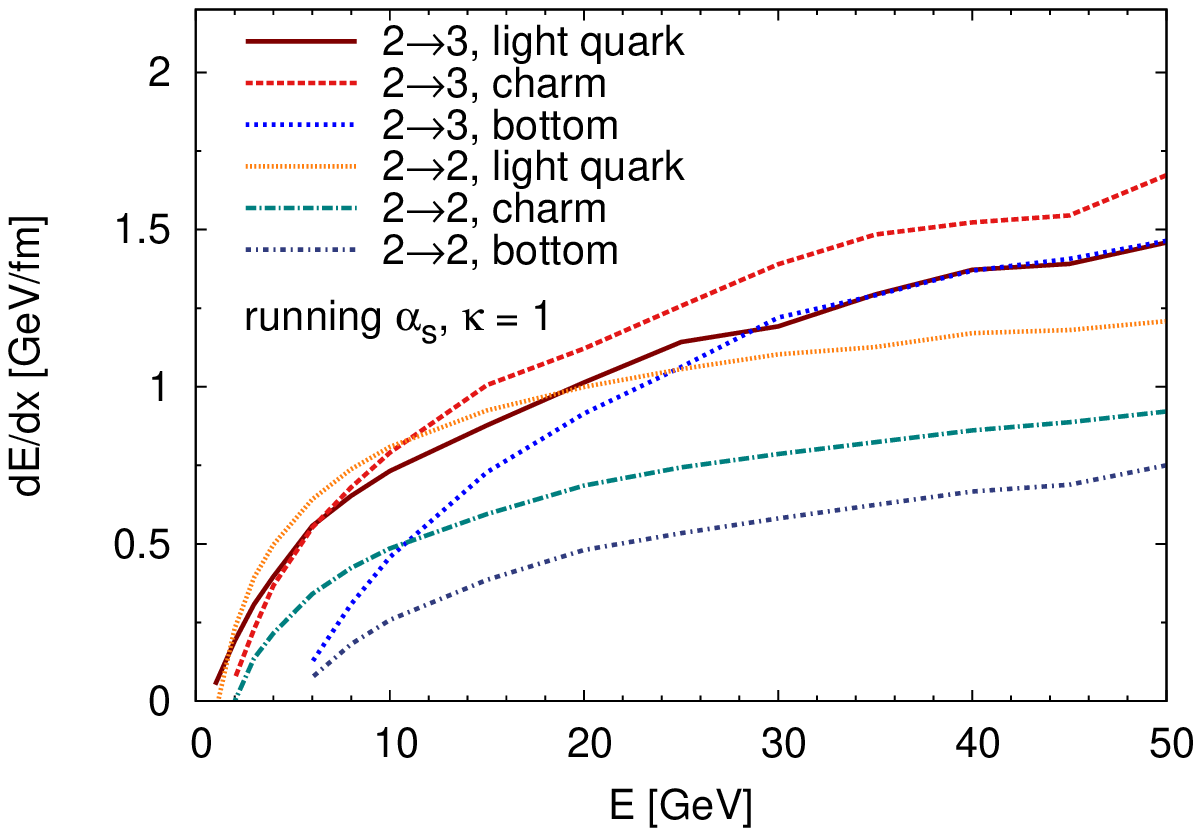}
\put(70,60){\footnotesize preliminary} 
\end{overpic}
\end{minipage}
\hfill
\begin{minipage}[t]{0.49\textwidth}
\centering
\begin{overpic}[width=1.0\textwidth]{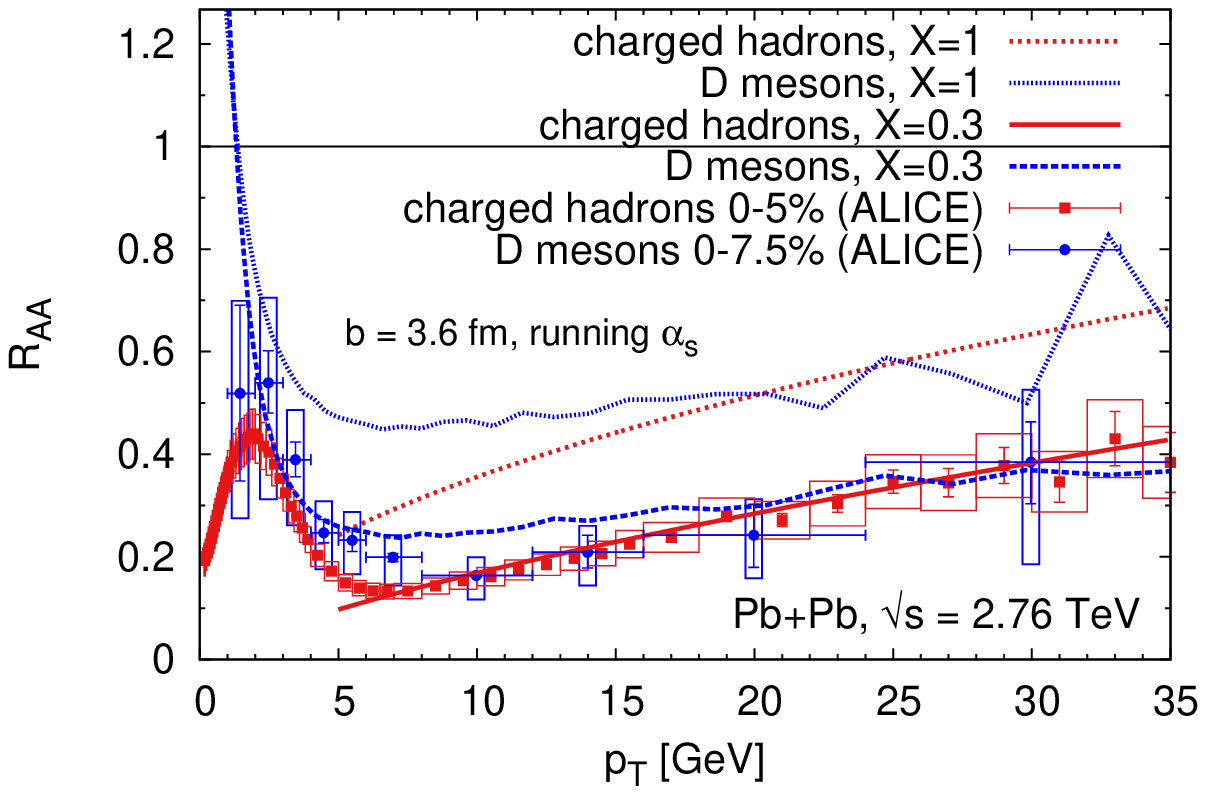}
\put(20,60){\footnotesize preliminary} 
\end{overpic}
\end{minipage}
\caption{Left: Elastic and radiative energy loss per unit length of a light quark ($M=0\unit{GeV}$), a charm quark ($M=1.3\unit{GeV}$), and a bottom quark ($M=4.6\unit{GeV}$) traversing a static thermal medium with temperature $T=0.4\unit{GeV}$. The curves are calculated with running coupling and Debye mass prefactor $\kappa = 1$. Right: Preliminary results on the nuclear modification factor $R_{AA}$ of charged hadrons and $D$ mesons at LHC in comparison to data \cite{Abelev:2012hxa,Grelli:2012yv}. The LPM parameter $X$ is set to 1 and 0.3.
}
\label{fig:dedx_raa_lhc}
\end{figure}

In the left panel of Fig.~\ref{fig:dedx_raa_lhc} the elastic and radiative energy loss of light and heavy quarks in a static thermal medium is depicted.
The radiative energy loss is on the same order or only slightly larger than the the elastic energy loss for all quark masses. Although the mass hierarchy is visible for the elastic energy loss, the radiative energy loss of all flavors is of the same size. The reason lies in the implementation of the Landau-Pomeranchuk-Migdal (LPM) effect producing a second dead cone at small emission angles that overshadows the dead cone due to the heavy quark mass.
This similar energy loss of light and charm quarks is indeed part of the explanation why the measured nuclear modification factors of charged hadrons and $D$ mesons  in heavy-ion collision have the same values. Furthermore, mass effects in the fragmentation of gluons and light quarks to charged hadrons and charm quarks to $D$ mesons lead to a very similar suppression of charged hadrons and $D$ mesons in BAMPS, as is depicted in the right panel of Fig.~\ref{fig:dedx_raa_lhc}.

Although the shape of the nuclear modification factor as a function of the transverse momentum is nicely reproduced by BAMPS (see $X=1$ curve in the right panel of Fig.~\ref{fig:dedx_raa_lhc}), the overall suppression is underestimated. The reason for this discrepancy is probably the effective implementation of the LPM effect in BAMPS, which discards all possible interference effects and only allows independent scatterings. If we introduce a factor $X<1$ that modifies the LPM cut-off and  effectively allows more radiative interactions, a good agreement with the experimental data of charged hadrons and $D$ mesons is found for $X=0.3$. Although the exact value of $X$ is a free parameter, we expect that a more sophisticated implementation of the LPM effect would effectively correspond to an $X<1$ and might make the need of the $X$ parameter obsolete.

\section{Summary}

We presented heavy flavor calculations with the parton cascade BAMPS. Allowing only elastic interactions with a running coupling and an improved Debye screening, the experimental data of all heavy flavor particles at LHC can be described if the binary cross sections are multiplied with $K=3.5$, which is tuned to the elliptic flow data at RHIC.
Furthermore, we show first results including also radiative processes. To this end, the elastic and radiative energy loss of light and heavy quarks in a static medium is shown. Furthermore, we compare BAMPS results to the experimentally measured nuclear modification factor of charged hadrons and $D$ mesons at LHC.
It would be interesting for a future project to study the impact of radiative processes on heavy flavor correlations \cite{Uphoff:2013rka} or heavy flavor tagged jets \cite{Senzel:2013dta}.

\section*{Acknowledgements}
This work was supported by the Bundesministerium f\"ur Bildung und Forschung (BMBF), the NSFC under grant No.\ 11275103, HGS-HIRe, and the Helmholtz International Center for FAIR within the framework of the LOEWE program launched by the State of Hesse. Numerical computations have been performed at the Center for Scientific Computing (CSC).

\section*{References}
\bibliography{hq,text}

\end{document}